# A Theory of Decomposition of Complex Chemical Networks using the Hill Functions

Eisuke Chikayama[1,2,3,*] and R. Craig Everroad[4]
*1 Niigata University of International and Information Studies, Mizukino 3-1-1, Nishi-ku, Niigata 950-2292, Japan*
*2 Environmental Metabolic Analysis Research Team, RIKEN, 1-7-22 Suehiro, Tsurumi-ku, Yokohama-shi, Kanagawa 230-0045, Japan*
*3 Image Processing Research Team, RIKEN, 2-1 Hirosawa, Wako-shi, Saitama 351-0198, Japan*
*4 Biosphere Oriented Biology Research Unit, RIKEN, 2-1 Hirosawa, Wako-shi, Saitama 351-0198, Japan*
* Author to whom correspondence should be addressed. Tel.: +81-25-239-3706 Fax: +81-25-239-3690 E-mail: chikaya@nuis.ac.jp

**The design and synthesis of complex and large mimicked biochemical networks de novo is an unsolved problem in synthetic biology. To address this limitation without resorting to *ad hoc* computations and experiments, a predictive mathematical theory is required to reduce these complex chemical networks into natural physico-chemical expressions. Here we provide a theory that offers a physico-chemical expression for a large chemical network that is almost arbitrarily both nonlinear and complex. Unexpectedly, the theory demonstrates that such networks can be decomposed into reactions based solely on the Hill equation, a simple chemical logic gate. This theory, analogous to implemented electrical logic gates or functional algorithms in a computer, is proposed for implementing regulated sequences of functional chemical reactions, such as mimicked genes, transcriptional regulation, signal transduction, protein interaction, and metabolic networks, into an artificial designed chemical network.**

## INTRODUCTION

A living cell is one of if not the most complex functional chemical networks known to exist. This network consists of biomolecules including genes, transcripts, proteins, polysaccharides, lipids, and metabolites interacting in a solution of water and ions. The emulation of such a system through *de novo* synthesis of a synthetic biochemical network that can mimic a living cell or complex cellular processes is one of the main goals of synthetic biology (Benner and Sismour, 2005; Elowitz and Lim, 2010; Khalil and Collins, 2010; Kwok, 2010; Purnick and Weiss, 2009; Way et al., 2014). The successful development of such a system, a completely designed and constructed artificial cell, will be a major scientific accomplishment. There have already been many successful examples in both theoretical (Alon, 2007; Karr et al., 2012; Savageau, 1969; Tomita et al., 1999) and experimental advancements (Alon, 2007; Annaluru et al., 2014; Becskei and Serrano, 2000; Deans et al., 2007; Elowitz and Leibler, 2000; Gardner et al., 2000; Gibson et al., 2010) in which whole-cell computation (Karr et al., 2012) and the synthetic genome (Annaluru et al., 2014; Gibson et al., 2010) have been rigorously pursued towards this accomplishment.

Savageau's Biochemical Systems Theory (BST) (Savageau, 1969) describes a large biochemical network in a unified way. It includes mathematical models, the S-system and the generalized mass action (GMA) system. These models have been successfully applied to many studies with their advantages in simplicity and generality (Alvarez-Vasquez et al., 2005; Savageau, 1976; Voit and Radivoyevitch, 2000). Specifically, the BST is the only theory, to our knowledge, that has been mathematically demonstrated to be able to describe almost arbitrarily nonlinear chemical systems, using the so-called canonical form (Savageau and Voit, 1987). However, there are also some known limitations such as the concept of operating points (Savageau, 1969) or regenerated artificial variables in the canonical forms. Thus there has never been a proposed mathematical theory for describing an arbitrarily nonlinear, and complex chemical network that can always be naturally interpreted with a physico-chemical expression. This means that synthetic biologists are not guaranteed to be able to synthesize an arbitrarily desired *de novo* large working biochemical network, which will be nonlinear and complex. Therefore, the development of a unified, holistic and predictive theory capable of dissecting an arbitrary, large, nonlinear, and complex chemical network into a set of natural physico-chemical expressions is anticipated.

Recently, a Dirac's formula in quantum mechanics (Dirac, 1958) with the famous Dirac delta function $\delta$ for an arbitrary function $R$,

$$R(0) = \int_{-\infty}^{\infty} R(\mu)\delta(\mu)d\mu$$
$$= R(-\infty) + \int_{-\infty}^{\infty} \frac{dR(\mu)}{d\mu}\sigma(-\mu)d\mu\,, \quad \text{(Eq. 1)}$$

is extended to that for multiple variables (Chikayama, 2014). It states that

$$R(X_1, X_2, \cdots, X_N) = \sum_{\vec{m}\in\{0,1\}^N} \int_0^{\infty} \cdots \int_0^{\infty} \frac{\partial^m R(m_1\mu_1, \cdots, m_N\mu_N)}{\partial\mu_1^{m_1} \cdots \partial\mu_N^{m_N}} \prod_{i=1}^{N} \{\sigma(X_i - \mu_i)d\mu_i\}^{m_i}\,, \quad \text{(Eq. 2)}$$

where $\sigma$ is the Heaviside step function (; $\vec{m} = (m_1, \cdots, m_N)$, $m = m_1 + \cdots + m_N$, and $m_i = 0$ or 1.) We found that this mathematical formula can be applied to a theory of the theoretical design of complex chemical networks. The theory we describe here concludes that each of all the chemical reactions included in chemical networks that are almost arbitrarily both nonlinear and complex can be synthesized based solely on the Hill equation (Hill, 1910), a simple chemical logic gate.

## RESULTS AND DISCUSSION

### Decomposition of complex chemical networks

Here we propose a mathematical theory based on the Hill equation (Hill function) (Hill, 1910), a simple chemical logic gate, for helping in the design of the theoretical and experimental synthesis of a *de novo* large working chemical network that is almost arbitrarily nonlinear and complex (Fig.

1.). In a general sense, this theory states that a mixture of chemical reactions, each described by the Hill functions, can be used to describe any set of interrelated chemical processes. More specifically, this theory is embodied in the following mathematical statement:

*Any differential equations for describing a chemical network constituting N substances $\{X_1, \cdots, X_N\}$ as shown here*:

$$\begin{cases} \dfrac{dX_1}{dt} = R_1(\{X_1,\cdots,X_N\}) \\ = \displaystyle\sum_{\vec{m}\in\{0,1\}^N} \int_0^\infty \cdots \int_0^\infty \frac{\partial^m R_1(m_1\mu_1,\cdots,m_N\mu_N)}{\partial\mu_1^{m_1}\cdots\partial\mu_N^{m_N}} \prod_{i=1}^{N} \{\sigma(X_i-\mu_i)d\mu_i\}^{m_i} \\ \vdots \\ \dfrac{dX_N}{dt} = R_N(\{X_1,\cdots,X_N\}) \\ = \displaystyle\sum_{\vec{m}\in\{0,1\}^N} \int_0^\infty \cdots \int_0^\infty \frac{\partial^m R_N(m_1\mu_1,\cdots,m_N\mu_N)}{\partial\mu_1^{m_1}\cdots\partial\mu_N^{m_N}} \prod_{i=1}^{N} \{\sigma(X_i-\mu_i)d\mu_i\}^{m_i} \end{cases}$$

(Eq. 3)

*provided that $R_i(\{0,\cdots,0\}) = 0$ can be approximated with any level of precision by*:

$$\begin{cases} \dfrac{dX_1}{dt} = \displaystyle\sum_{j=1}^{n_1} \alpha_{1j} s_{1j} \\ \vdots \\ \dfrac{dX_N}{dt} = \displaystyle\sum_{j=1}^{n_N} \alpha_{Nj} s_{Nj} \end{cases},$$

(Eq. 4)

*in which an elementary nonlinear function*:

$$s_{ij}(\{X_1,\cdots,X_N\}) = \prod_{k=1}^{N} \frac{X_k^{\beta_{ijk}}}{X_k^{\beta_{ijk}} + \mu_{ijk}^{\beta_{ijk}}},$$

(Eq. 5)

*is a direct product of the Hill functions* ($X_i \geq 0$, $\alpha_{ij} \geq 0, \beta_{ijk} \geq 0$, *and* $\mu_{ijk} \geq 0$).

(Eq. 3)

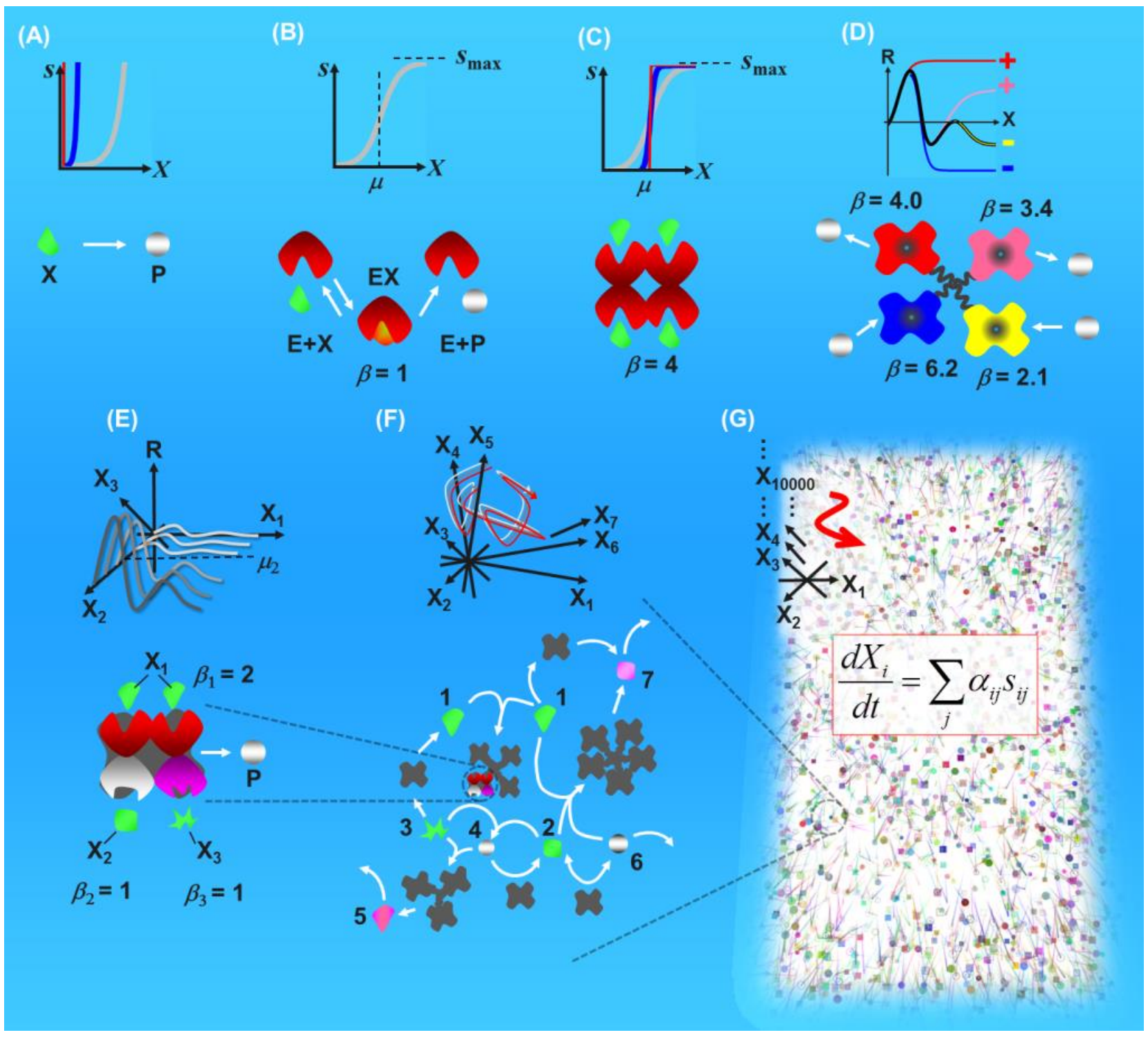

**Figure 1. Physico-chemical interpretation of the theory.** It concludes that almost any complex chemical networks can be artificially synthesized in principle. The theory guarantees superposition of the Hill functions, the mixture of reactive chemical substances in sigmoidal reactions can be an almost arbitrarily nonlinear and complex chemical network. Note that all the substrate axes *X* in the figure are logarithmic. (A) Simple chemical reaction rates (*s*) such as a product from a substrate (*X*). The rates can diverge. (B) A simple sigmoidal reaction rate described by the Hill function with $\beta = 1$. It has a saturated rate ($s_{max}$). A sigmoidal reaction is regarded as a chemical logic gate. It gradually switches on or off, depending on whether the reaction rate is greater or less than the substrate threshold $\mu$, respectively. A chemical logic gate can transfer chemical information. An example of the chemical logic gate is a Michaelis-Menten reaction (below), of which the Hill coefficient $\beta$ is 1. Note that the substrate axis is logarithmic and thus the curve is sigmoidal, which is not usual for drawing the Michaelis-Menten reaction. (C) Sigmoidal reaction rates can be steep with higher Hill coefficients. A steep reaction rate with $\beta = 4$ (blue) and Michaelis-Menten rate with $\beta = 1$ (light gray) are shown. Ideally, the rate becomes a discrete digital switch (the Heaviside step function, mathematically,) with $\beta$ = infinity (red line). An example of the steep chemical logic gate is a tetrameric substrate-enzyme complex that produces a product only after the same four substrates bind (below) in which $\beta = 4$ ideally. (D) An example of a more intricate nonlinear rate. It can be the sum of four different sigmoidal reaction rates (upper, red, blue, pink, and yellow lines). The superposition can be physico-chemically interpreted by a supramolecular catalyst (lower). It consists of four loosely linked molecular complexes (red, blue, pink, and yellow multiple enzymes, lower) in which each complex has a different Hill coefficient and direction for the chemical reaction (white arrows, lower; + and –, next to the colored lines, upper) but the same product. (E) An example of a heterotetrameric complex that binds two $X_1$, one $X_2$, and one $X_3$ substrates (lower) respectively. Each of the binding sites has $\beta_1 = 2$, $\beta_2 = 1$, or $\beta_3 = 1$ and the heterotetramer produces a product only after binding all the substrates. It corresponds to a direct product with *N* of 3 in Eq. 5. The shape of the nonlinear curve in (D) is supposed to be kept in the $X_1$ axis (upper) but it is totally on or off depending on $X_2$ with a threshold $\mu_2$ (dotted line in the $X_1$-$X_2$ plane, upper), which corresponds to the direct product between $X_1$ and $X_2$ terms in Eq. 5. (F) An example of an imaginary synthesized chemical network composed of 7 substrates with fixed catalysts (lower). This system can be described by the unified mathematical expression, Eqs. 4 and 5, in the same manner. The design of the chemical network can be achieved by drawing orbits and rates at all the points in the 7-dimensional substrate concentration space (upper, red and white arrows are examples of orbits). Only one rate should exist at any point in the space (This property is called *autonomous*). This drawing can be converted to Eq. 3 and then to the real constants in Eqs. 4 and 5 that are guaranteed to exist by the theory. The constants are supposed to be acquired in the chemical network (lower) in the dissociation constants, Hill coefficients, and saturated rates. (G) Further interpretation makes possible the synthesis of a large chemical network that is almost arbitrarily nonlinear and complex. It can acquire any number of components (upper left, e.g., $X_{10000}$), any orbits and rates (red arrow), and diverse chemical reactions (background) based on an expression Eq. 4 (center). The Eq. 4 is the linear combination of Eq. 5, a direct product of the Hill functions. The Eq. 5 is a nonlinear and basic chemical logic gate.

The theory is simply derived by replacing, in Eq. 3, derivatives with real constants, integrals with summations, and the Heaviside step functions with the Hill functions, respectively, since the derivatives are constant real values at the specific values of $\mu$, a summation approximates to integrals, and the Hill function approximates to the Heaviside step function. The importance of this theory is not the replacement of mathematics but is the finding of a unified, holistic and predictive theory capable of dissecting an arbitrary, large, nonlinear, and complex chemical network into a set of phenomenological physico-chemical expressions, which first gives synthetic biologists the theoretical basis of syntheses of arbitrarily desired *de novo* large working mimicked chemical networks, which will be nonlinear and complex.

The theory shows that Eq. 3, which can phenomenologically represent almost any chemical system, is equivalent to Eqs. 4 and 5, which represent an artificial synthesized chemical system described by the Hill functions (the word "almost" means that the right-hand sides of Eq. 3 are a type of derivable and continuous functions. The definition is strictly described elsewhere (Chikayama, 2014).) The linear combinations in Eq. 4 are simply the mixture of the chemical substances or supramolecular catalysts representing the artificially synthesized chemical system.

**Synthetic biologist can synthesize arbitrarily mimicked chemical networks**

The theory is to our knowledge the first mathematical theory that in principle guarantees the synthesis of a chemical network that is almost arbitrarily nonlinear and complex based solely on physico-chemically interpretable equations. Moreover, because arbitrarily nonlinear chemical networks can implement arbitrarily logical algorithms (logically designed sequences of chemical reactions), it implies that biochemical networks in living organisms can also implement almost any functional algorithm (analogous to a computer being able to implement almost any functional programs), at least for a spatially-homogeneous mixture in a local region inside a cell. The similar point was proposed by Bray using examples of aspartate transcarbamoylase, calcium/calmodulin-dependent protein kinase II, glycogen synthase, or a protein circuit mediating chemotactic responses of coliform bacteria (Bray, 1995). The concept is recently summarized as *a computer in every living cell* (Bray, 2009).

**Theoretical design of complex chemical networks**

The theory includes the concepts to allow for the design of an almost arbitrarily nonlinear chemical network. With this theory, a complex chemical network can be dissected into a set of physico-chemical expressions using only the equivalent concepts: the Hill function and chemical logic gate. Figure 1 details this principle conceptually. In a simple chemical reaction a substrate changes to a product with the reaction rate monotonically dependent on the concentration of the substrate (Fig. 1A), whereas the Hill function exhibits strong nonlinearity, i.e., a sigmoidal reaction (Figs. 1B and C. Note that all the substrate axes $X$ in the figure are logarithmic). The sigmoidal reaction in a chemical system can be interpreted as an analog switch (Daniel et al., 2013) or as a basic chemical logic gate. In the Hill function, all the constants $\alpha_{ij}$, $\mu_{ijk}$, and $\beta_{ijk}$ in Eqs. 4 and 5 represent physico-chemically meaningful phenomenological parameters, i.e., a saturated reaction rate, dissociation constant, and cooperativity (or Hill coefficient). A Hill coefficient ideally becomes $n$ in the reaction $E + n \times X \leftrightarrow EY \rightarrow E + P$ in which an enzyme $E$ and $n$ molecules of substrate $X$ produce a product $P$ through an enzyme-$n$-substrates complex $EY$ (Fig. 1B ($n$ = 1) and Fig. 1C ($n$ = 4)). It is well known, however, that in practice a Hill coefficient is a fractional (not only integer but also real) value (Weiss, 1997), which is preferable for our theory. In general, larger values of Hill coefficients indicate higher cooperativity, or more highly digitized behaviors for information processing (Shinohara et al., 2014). In this way Eqs. 4 and 5 can describe a more intricate nonlinearity. For example, suppose there exists a reaction with a net rate $R$ that exhibits a strongly nonlinear dependence upon the substrate concentration $X$ (Fig. 1D, upper, black solid line) described by the right-hand side of Eq. 3, $R(X)$. From the theory, we can decompose it into a linear combination of the Hill functions using Eqs. 4 and 5 (Fig. 1D, upper, red, blue, pink, and yellow solid lines). In sum, these colored lines are equal to the black solid line that represents the net reaction rate $R(X)$. Additionally those colored lines are all logic gates (individual examples shown in Fig. 1B and C, lower) that can be physico-chemically (mathematically) interpreted as a molecular complex of cooperative enzymes (logic gates), that result in the net rate and, for example, together act as a supramolecular catalyst (Fig. 1D, lower). The superimposition of these four molecular complexes results in the complicated net rate reaction $R(X)$ (Fig. 1D, upper, black solid line). Further, a multidimensional (direct product) form of the Hill functions, Eq. 5, represents simultaneous or sequential bindings of multiple distinct substrates or ligands (Fig. 1E, bottom, green objects labeled $X_1$-$X_3$) before enzymatic activity (Cornish-Bowden, 2004) (Fig. 1E, $P$, white). In this case, the axis $X$ as shown in Fig. 1D (upper) is extended to three axes $X_1$-$X_3$ in Fig. 1E (upper). Such an interpretation can easily be extended even further to a more complicated system containing supramolecular catalysts, multiple enzymes, substrates, and ligands like a circuit consisting of many wired chemical logic gates (Fig. 1F), all under the same mathematical expressions given in Eqs. 4 and 5. In this latter complex case (Fig. 1F), whereas the axes are extended to $X_1$-$X_7$, there are many of multiple reaction rates and thus the axis $R$ as in Fig. 1E is avoided, which is a standard way of describing a dynamical system.

The design of a complex chemical network using this theory is performed as an inverse problem. If a designer wants to predict how a chemical network will behave from given initial conditions, they should draw orbits in a coordinate space (specifically, state or phase space) from their starting points (e.g., Fig. 1F, upper, red and white orbits). A point in the phase space represents a set of specific concentrations of all the constituent chemical substances. Thus the number of dimensions in the phase space is equal to be the number of substances $\{X_1, \cdots, X_N\}$ being modeled. The multiple starting points correspond to multiple initial conditions. The orbits correspond to the desired dynamics in the concentrations of the constituents. Once all orbits are drawn, a point on any given orbit uniquely determines the concentrations of all the substances $\{X_1, \cdots, X_N\}$ in the system. Additionally, the designer should determine the rate of the instant direction of the orbit at that point. It corresponds to give a rate vector at that point. In this manner, designing desired orbits and then determining the rate vector at each of the points on all the orbits in the phase space are equal to drawing all the orbits and rate vectors in the phase space in bi-infinite time (specifically, the set of rate vectors is a vector field in the phase space). Note that the designer can choose a favorable length of the rate vector at the point on an orbit keeping the instant direction, and all the designed orbits cannot cross each other. All the specific drawn orbits and rate vectors totally correspond to a set of specific nonlinear functions, and thus can be expressed by Eq. 3. Using the theory, it can be converted to Eqs. 4 and 5, which have specific parameters. The set of Eqs. 4 and 5 is a mixture of reactively interrelated chemical constituents described by sigmoidal reactions, i.e., functionally wired chemical logic gates. This is the design of a chemical network using the theory. Specifically, the design of all orbits and rate vectors in the phase space corresponds to the unique determination of the right-hand side of Eq. 3. Because Eqs. 4 and 5 approximate to Eq. 3, the set of Eqs. 4 and 5 that have specific parameters is the blueprint of the chemical network. It is noteworthy that the theory does not limit the number of components in a given system; consequently a large chemical network that is almost arbitrarily nonlinear and complex can be designed at will (Fig. 1G) by using Eqs. 4 and 5, in principle. Since Eq. 4 is described without Cartesian space coordinates, an easy extension of Eq. 4 may be used to include elements of diffusive factors $D_i$ into a system,

$$\begin{cases} \dfrac{\partial X_1}{dt} = \sum_{j_1} \alpha_{1j_1} s_{1j_1} + D_1 \nabla^2 X_1 \\ \qquad\qquad \vdots \\ \dfrac{\partial X_N}{dt} = \sum_{j_N} \alpha_{Nj_N} s_{Nj_N} + D_N \nabla^2 X_N \end{cases}, \quad \text{(Eq. 6)}$$

though this extension does not guarantee spatially designed arbitrary chemical networks.

The Hill equation was established in 1910 by A. V. Hill to describe a sigmoidal reaction on saturated hemoglobins by dioxygen (Hill, 1910) and well-known examples of such sigmoidal reactions in living systems are the rate of mRNA production vs. the concentration of transcription factors (Ackers et al., 1982; Gardner et al., 2000; Yagil and Yagil, 1971), the rate for certain types of allosteric enzymes vs. substrate concentration (Hammes and Wu, 1974), tension in cardiac muscle vs. noradrenaline concentration (Reuter, 1974), and the firing rate of a neuron vs. synaptic input potential (McCulloch and Pitts, 1943; Silver, 2010). Protein circuits (Bray, 1995) and transcriptional network motifs (Alon, 2007) are examples of critical components for maintaining living functions, in which numbers of sigmoidal reactions are wired and play essential roles for chemical information transfer. The successes of implementing artificially synthesized genetic circuits into bacteria (Becskei and Serrano, 2000; Elowitz and Leibler, 2000; Gardner et al., 2000) and mammalian cells (Deans et al., 2007; Tigges et al., 2009) tell us the sigmoidal reactions are deeply related to the design principles and the mechanisms of genetic networks. The formalization of a similar multidimensional description as in Eq. 5 has already had an experimental consensus in transcriptional regulation network studies (Setty et al., 2003) and mathematical models with a variety of the Hill functions or sigmoid functions. The use of analogues of Eqs. 4 and 5 has been a known empirical approach aimed at analyzing a large biochemical network such as a transcriptional regulatory or signal transduction network (Aldridge et al., 2006; de Jong, 2002); the fundamental basis of these empirical approaches can be given by this theory.

Practically the theory, as presented here, is not intended for describing the complexity of a real living system. Rather the intent is to promote the development of synthetic chemical networks that can mimic living systems. An example of the method for synthesized mimics of chemical reactions in living systems, where our theory is combined with Darwinian evolutionary theory, can be in principle used to evolve large (*in vitro* or *in silico*) chemical networks. Networks are initially (experimentally or computationally) synthesized as in Eqs. 4 and 5 with diverse constants $N$, $\alpha_{ij}$, $\mu_{ijk}$, and $\beta_{ijk}$; and can then be 'evolved' through mimicked phenotypic effects of mutation (by perturbing or altering the desired constants) and natural selection (by selecting the best 'mutants' by some objective criteria) towards the 'fitter' constants. This can be done repeatedly. Ultimately, the experimental synthesis of complex chemical networks that incorporate tremendous numbers of chemical molecules as mimics of functions in living systems through theoretical design will offer important clues in the quest of understanding 'what life is' in physico-chemical language (Schrödinger, 1944).

**Author contribution**

EC devised the theory. EC and RCE co-wrote the manuscript.

**Acknowledgements**

We thank N. Nakamichi, K. Kihara, S. Moriya, H. Yokota, J. Kikuchi, and K. Soda for discussion.